\documentclass{PoS}

      \newcommand{\pp}{\mbox{pp}\xspace}
      
      \newcommand{\ppb}{\mbox{p--Pb}\xspace}
      \newcommand{\pb}{\mbox{Pb--Pb}\xspace}

  \newcommand{\krl}{\ensuremath{\kern-0.18em}}
  \newcommand{\krr}{\ensuremath{\kern-0.09em}}
  \newcommand{\tms}{\ensuremath{\kern-0.1em\times\kern-0.2em}}
      
      \newcommand{\dd}{\ensuremath{\mathrm{d}}}

      \newcommand{\dndykrl}{\ensuremath{\dd N\krl/\krr\dd y}\xspace}
      \newcommand{\pt}{\ensuremath{p_{\mathrm{T}}}\xspace}

      \newcommand{\ptopi}{\ensuremath{({\rm p+\bar{p}}) / (\pi^{+}+\pi^{-})}\xspace}
      \newcommand{\ptopisum}{\ensuremath{{\rm p } / \pi}\xspace}
      \newcommand{\ktopi}{\ensuremath{({\rm K}^{+}+{\rm K}^{-}) / (\pi^{+}+\pi^{-})}\xspace}
      \newcommand{\ktopisum}{\ensuremath{{\rm K } / \pi}\xspace}

      \newcommand{\omegatopisum}{\ensuremath{\Omega / \pi}\xspace}
      \newcommand{\xitopisum}{\ensuremath{\Xi / \pi}\xspace}

      \newcommand{\pix}{\ensuremath{\pi^{\pm}}\xspace}

      \newcommand{\kx}{\ensuremath{\mathrm{K}^{\pm}}\xspace}

      \newcommand{\prx}{\ensuremath{{\rm p}(\rm \bar{p})}\xspace}

      \newcommand{\chpi}{\ensuremath{\pi^{+}+\pi^{-}}\xspace}

      \newcommand{\lap}{\ensuremath{\Lambda}\xspace}
      \newcommand{\lam}{\ensuremath{\bar{\Lambda}}\xspace}

      \newcommand{\xim}{\ensuremath{\Xi^{-}}\xspace}
      \newcommand{\xip}{\ensuremath{\overline{\Xi}^{+}}\xspace}      
      
      \newcommand{\omm}{\ensuremath{\Omega^{-}}\xspace}
      \newcommand{\omp}{\ensuremath{\overline{\Omega}^{+}}\xspace}
      
      \newcommand{\kzs}{\ensuremath{\mathrm{K_{S}^{0}}}\xspace}      

      \newcommand{\ks}{\ensuremath{\mathrm{K^{*0}}}\xspace}
      \newcommand{\ksb}{\ensuremath{\mathrm{\overline{K}^{*0}}}\xspace}

      \newcommand{\ph}{\ensuremath{\phi}\xspace}

      \newcommand{\py}{\textsc{Pythia}\xspace}

      \newcommand{\dipsy}{\textsc{Dipsy}\xspace}
      \newcommand{\geant}{\textsc{Geant}\xspace}

      \newcommand{\eplhc}{\textsc{Epos-LHC}\xspace}

\usepackage[lofdepth=1,lotdepth]{subfig}
\usepackage{xspace}

\title{New results on the multiplicity and centre-of-mass energy dependence of identified particle production in \pp collisions with ALICE}

\ShortTitle{New results on identified particle production in \pp collisions with ALICE}

\author{\speaker{Gyula Benc\'edi}\\
        On behalf of the ALICE Collaboration\\
        E-mail: \email{bencedi.gyula@wigner.mta.hu}}

\abstract{

  The study of identified particle production in proton-proton (\pp) collisions as a function of center-of-mass energy ($\sqrt{s}$) and event charged-particle multiplicity is a key tool for understanding similarities and differences between small and large collision systems. 
  We report on new measurements of the production of identified particles and their dependence on multiplicity and $\sqrt{s}$. \\
  Latest results for light flavor hadrons, comprising \pix, \kx, \prx, single-strange (\kzs, \lam, and \lap), multi-strange (\xim, \xip, \omm, and \omp) particles as well as resonances (\ks, \ksb, \ph), are presented for $\sqrt{s}~=~5.02, 7$, and 13\,TeV\,---\,measurements for 
  $\sqrt{s}~=~5.02$\,TeV \pp collisions are reported here for the first time. 
  The measured minimum bias \pt spectra and yields were complemented with multiplicity-dependent measurements for single- and multi-strange hadrons at $\sqrt{s}~=~7$ and $\sqrt{s}~=~13$\,TeV. Results are compared to measurements at lower collision energies as well as to those in proton-lead (\ppb) and 
  lead-lead (\pb) collisions, respectively at $\sqrt{s}_{\rm NN}~=~5.02$\,TeV and $\sqrt{s}_{\rm NN}~=~2.76$\,TeV. \\
  The results unveil intriguing similarities among collision systems at different center-of-mass energies. The production rates of strange hadrons are found to increase more than those of non-strange particles, showing an enhancement pattern with multiplicity which does not depend 
  on the collision energy. These yield ratios take values which are alike for small systems at comparable multiplicities, and show smooth evolution with multiplicity across all collision systems; they tend to approach those measured in \pb collisions.
  Although, the multiplicity dependence of spectral shapes can be qualitatively described by general-purpose Monte Carlo (MC) event generators, the evolution of integrated yield ratios is barely (or not) captured at all by MC model predictions.         
          }

\FullConference{EPS-HEP 2017, European Physical Society conference on High Energy Physics\\
		5-12 July 2017\\
		Venice, Italy}

\newsavebox{\measurebox}

\usepackage{lineno}

\begin{document}

 %
    
  \section{Introduction}

  Recent results on particle production at the LHC obtained in high-multiplicity proton-proton (\pp) and proton-lead (\ppb) collisions revealed phenomena which are similar to those seen in \pb collisions where they are attributed to bulk collective effects~\cite{Voloshin:2008dg,Bala:2016hlf}.
  Apart from the peculiar patterns seen in two-particle azimuthal correlations~\cite{ABELEV:2013wsa,Khachatryan:2016txc}, notably, some of these observations are the radial flow signals~\cite{Abelev:2013haa,Adam:2016dau} and the 
  strangeness enhancement~\cite{ALICE:2017jyt,Khachatryan:2016yru}.
  Strangeness enhancement with respect to minimum bias (MB) \pp collisions historically has been proposed as a signature of Quark-Gluon Plasma (QGP) formation~\cite{Rafelski:1982pu}. However, in a more modern view, the production of strange particles 
  is discussed together with other light flavor hadrons in the context of thermal statistical models and hydrodynamics. 
  The strange to non-strange production rate shows an increasing trend with multiplicity in \pp and \ppb collisions and saturates for most central \pb collision, in good 
  agreement with thermal model predictions.
  Here, changing initial state configuration, i.e. the colliding system (\pp, \ppb, \pb), does not seem to modify relative particle abundances provided that the particle multiplicities are comparable.

  On the top of the study performed in multiplicity classes, by changing the collision energy one might get further insight into the particle production mechanism in \pp collisions.
  For this purpose, a comprehensive study of identified particle production has been performed by ALICE using \pp data recorded during the LHC Run 2 (2015\,--\,2018).

 %
    
  \section{Analysis details}

  ALICE~\cite{Aamodt:2008zz} is a dedicated heavy-ion experiment at the LHC which is optimised to study the properties of the strongly interacting deconfined medium of quarks and gluons, the QGP
  created in ultra-relativistic heavy-ion collisions~\cite{Bass:1998vz}.
    
  Besides colliding heavy ions, ALICE provides important contributions to the LHC \pp physics program, which is complementary to other LHC experiments due to its capability to measure particle production down to very low 
  transverse momentum ($\pt\simeq 100$\,MeV$/c$), and due to its excellent particle identification (PID) capabilities in the central barrel region ($|\eta| < 0.9$)~\cite{Abelev:2014ffa}.
  
  The most recent measurements presented in these proceedings were obtained from the analysis of a data sample consisting of $\sim50$ M events of \pp collisions at the top LHC energy, $\sqrt{s}~=~13$\,TeV, recorded by ALICE in 2015. 
  The data were collected using a minimum bias trigger, which required a hit in both V0 scintillator arrays in coincidence with the arrival of proton bunches from both directions along the beam.
  Events containing more than one primary vertex within 10 cm along the beam axis from the interaction point were considered as pile-up or beam-induced background and are discarded from the analysis. Corrections are calculated from Monte Carlo (MC)
  simulations, using \py~8 (Monash 2013 tune)~\cite{Sjostrand:2007gs} as particle generator along with \geant~3~\cite{Brun:1987ma} for describing particle transport within the ALICE detector.
  
  Identification of light flavor charged hadrons (\pix, \kx, \prx) as well as short lived particles (\kzs, $\Lambda$, $\Xi$ and $\Omega$) and resonances (\ks, \ksb, and \ph) was performed with similar techniques to those applied 
  in earlier measurements~\cite{Adam:2015qaa,Adam:2017zbf} during LHC Run 1 (2009\,--\,2013).

  For the measurement at $\sqrt{s}~=~7$\,TeV, minimum bias events have been collected in 2010 having low pile-up and requiring at least one charged particle in the pseudorapidity interval $|\eta|<1$, corresponding to about 75\% of the total inelastic (INEL) cross-section.  
  Here, particle production is studied as a function of the event activity, meaning that the data sample is divided in event multiplicity classes. The selection in multiplicity classes is performed via the sum of the 
  signal amplitudes of V0 scintillator arrays located at forward (V0A) and backward (V0C) rapidity, commonly referred to as V0M.
  The average charged particle pseudorapidity density, $\left <\rm{d}N_{\rm{ch}}/\rm{d}\eta\right >_{|\eta|<0.5}$, is estimated within each multiplicity class by the average number of tracks in the central pseudorapidity region $|\eta|<0.5$, in order to avoid autocorrelation biases.

 %
    
  \section{Results and discussion}

\begin{figure}[htb!]
	  \centering
	    \subfloat[\label{fig:figPP13A}]{
	    \includegraphics[width=0.39\textwidth]{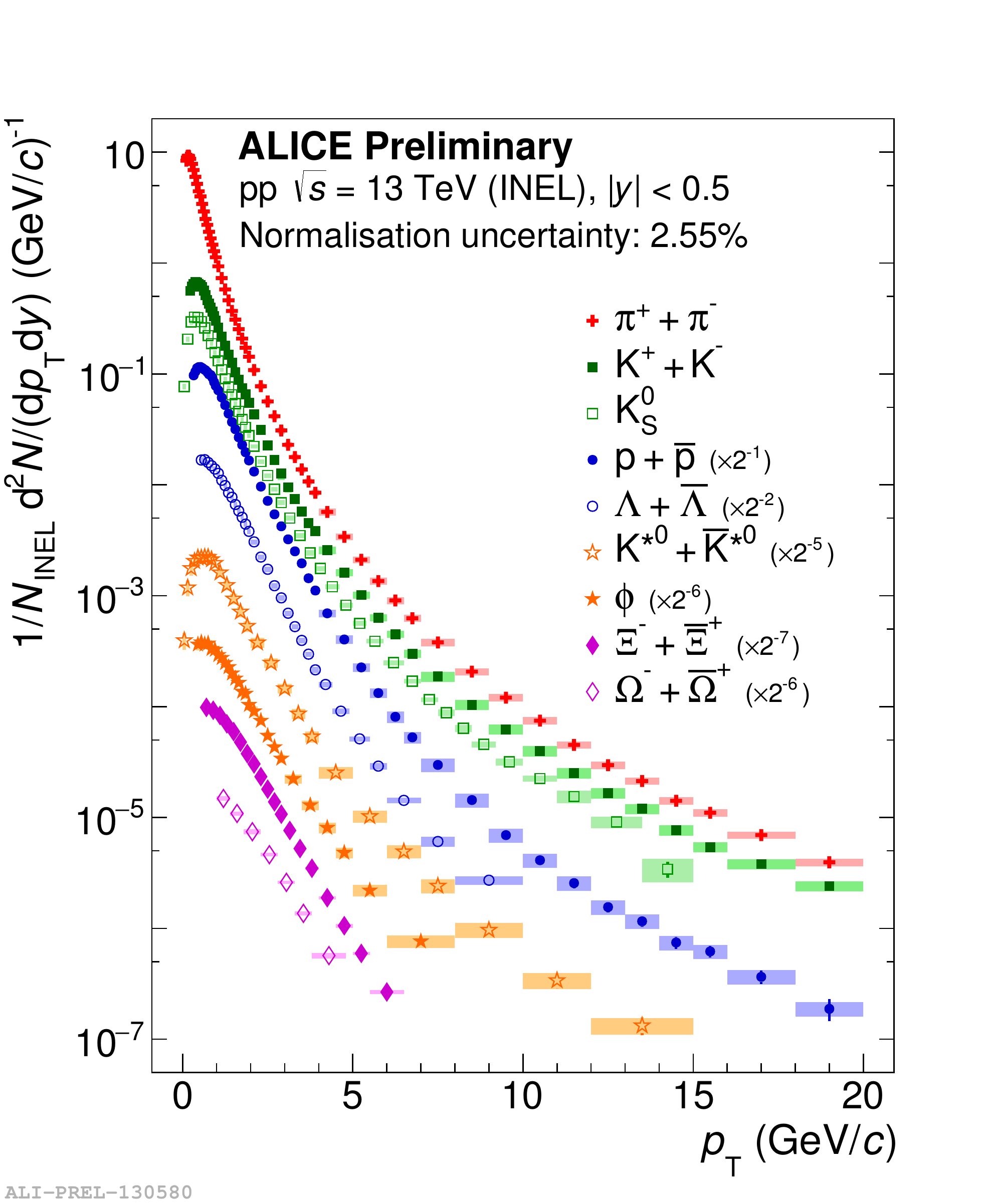}
	    }
	    \subfloat[\label{fig:figPP13B}]{
	    \includegraphics[width=0.36\textwidth]{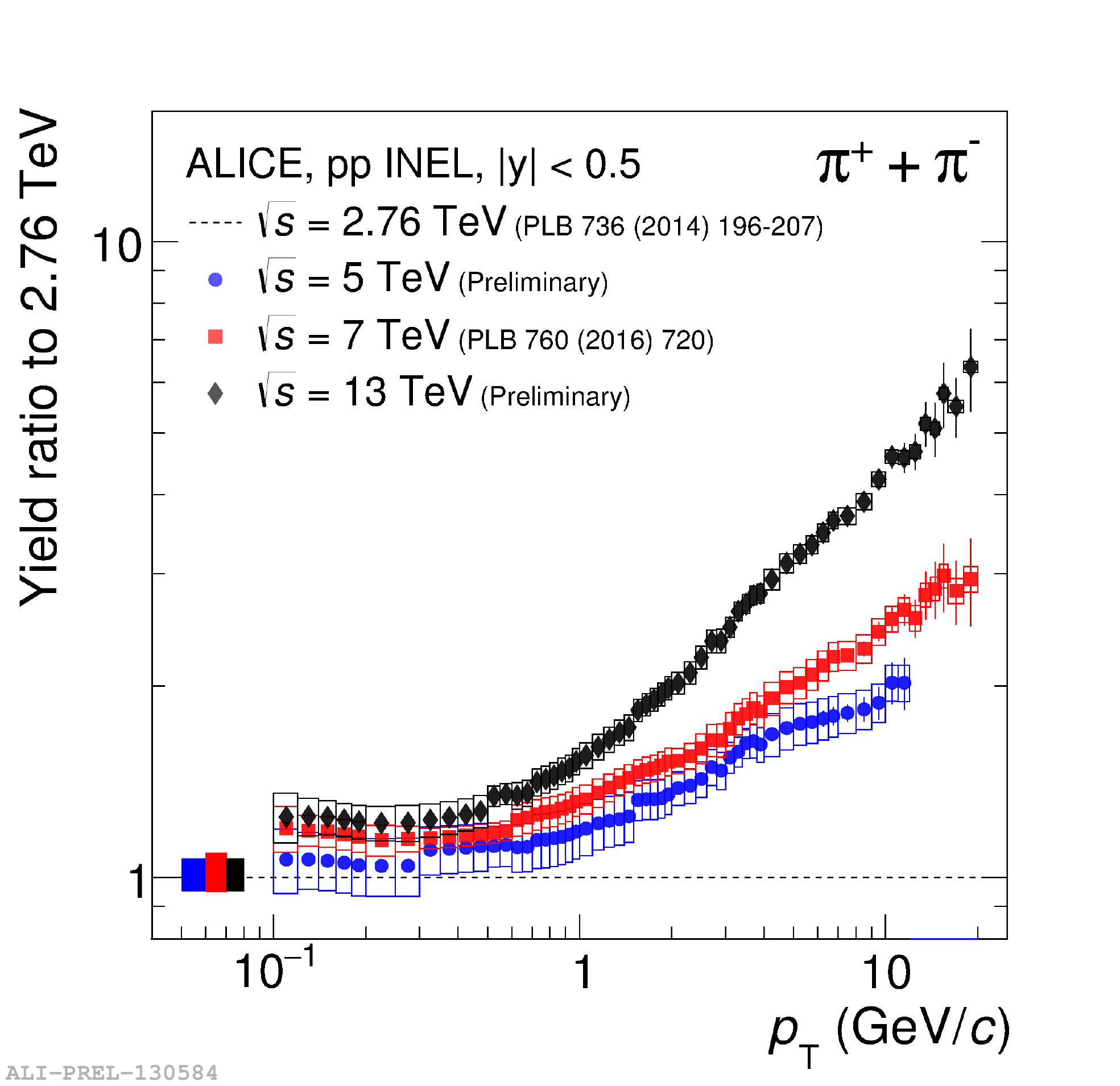}
	    }
	  \caption[]{\protect\subref{fig:figPP13A} \pt spectra of light flavor hadrons measured at mid-rapidity ($|y| < 0.5$) in minimum bias INEL \pp collisions at $\sqrt{s}~=~13$\,TeV.
	  \protect\subref{fig:figPP13B} Ratios of \pt spectra of charge-summed pions in inelastic events at various center-of-mass energies to that at $\sqrt{s}~=~2.76$\,TeV.
	  The statistical and systematic uncertainties are shown as vertical error bars and boxes, respectively. The normalisation uncertainties are indicated by boxes around unity.	  
	  }
	  \label{fig:figPP13}
	\end{figure}

	Figure~\ref{fig:figPP13A} presents a comprehensive collection of \pt distributions for light flavor hadrons measured in INEL \pp collisions at $\sqrt{s}~=~13$\,TeV. By comparing \pt spectra to those measured at $\sqrt{s}~=~2.76$\,TeV, a progressive hardening 
	of the spectral shapes with increasing $\sqrt{s}$ is observed for all particle species. Shown in Fig.~\ref{fig:figPP13B} is an example of the ratio of yields for charged pions where the latest measurements at $\sqrt{s}=5.02$ and $\sqrt{s}~=~13$\,TeV 
	are included. These ratios reveal two different \pt regimes: in the soft regime ($\pt\lesssim 1$\,GeV$/c$) neither the magnitudes nor the spectral shapes change significantly with \pt within uncertainties, whereas in the hard regime the opposite is observed, showing 
	a very significant dependence on $\sqrt{s}$.

\begin{figure}[htb!]
	  \centering
	  \includegraphics[keepaspectratio, width=0.55\columnwidth]{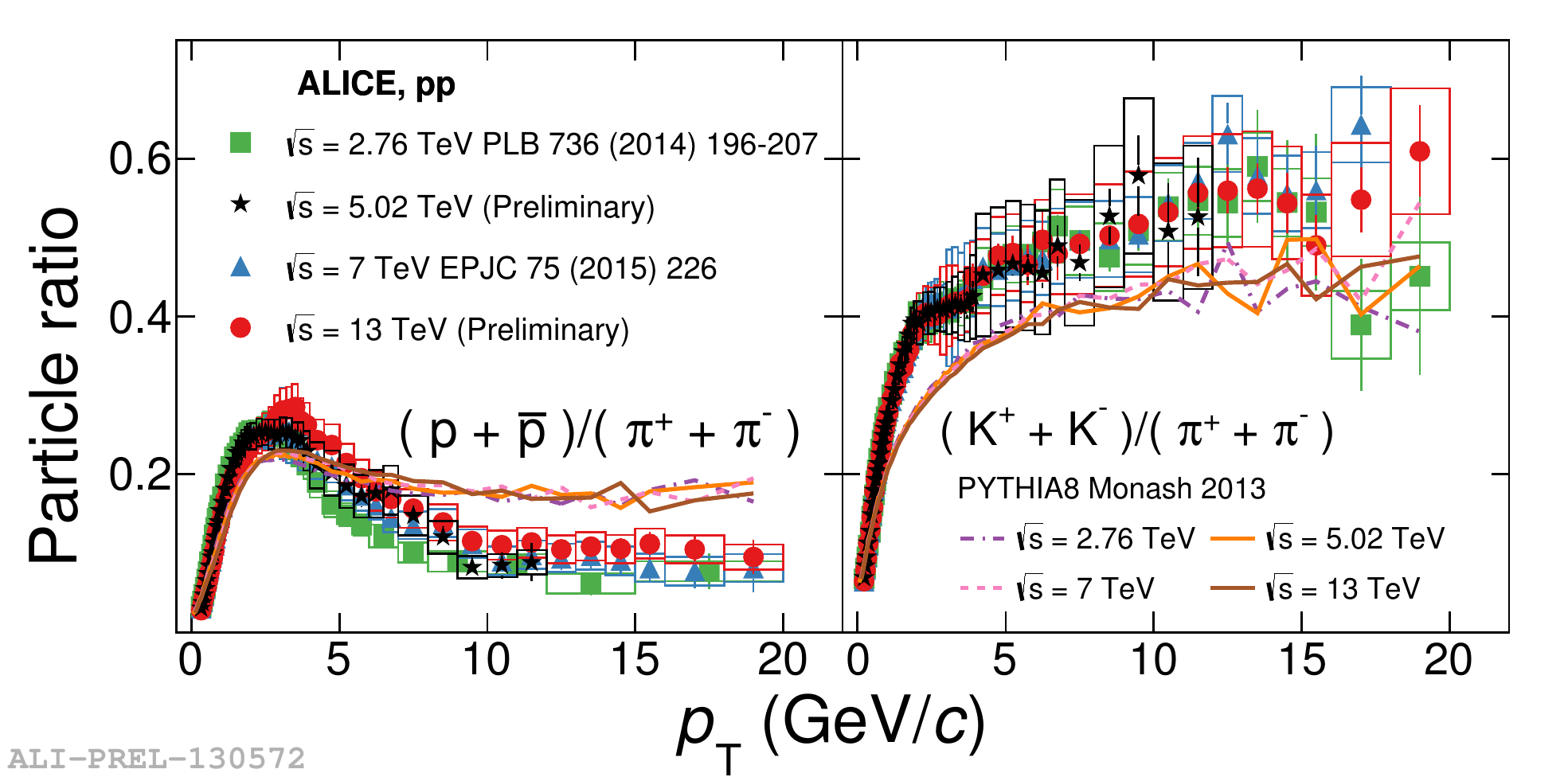}
	  \caption[]{Collision energy dependence of the \pt-differential \ptopi (left) and \ktopi (right) particle ratios measured 
	  in minimum bias \pp collisions, superimposed with \py~8 (Monash 2013) model predictions.
	  }
	  \label{fig:SpectraRatios}
	\end{figure}

	The \pt-differential \ptopisum and \ktopisum particle ratios are shown in Fig.~\ref{fig:SpectraRatios} measured in MB INEL \pp collisions at different $\sqrt{s}$, including \pp at $\sqrt{s}~=~5.02$\,TeV, reported here for the first time. 
	The baryon-to-meson (\ptopisum) ratios show a modest dependence with $\sqrt{s}$ in the intermediate \pt region, the position of the peak shifts towards higher \pt with increasing $\sqrt{s}$. No such behavior is present for \ktopisum. 
	For high \pt ($>10$\,GeV$/c$), no evidence of evolution with $\sqrt{s}$ is seen for any of the ratios within uncertainties. It is worth noticing that a minor modification of the \ptopisum ratio might be expected due to the moderate increase of $\left <\rm{d}N_{\rm{ch}}/\rm{d}\eta\right >_{|\eta|<0.5}$ 
	with $\sqrt{s}$~\cite{Adam:2015pza}.

	
      Predictions obtained from the \py~8 Monte Carlo generator are shown, which are unable to describe simultaneously both ratios in the entire \pt range. Also, the $\sqrt{s}$ evolution of the ratios is not reproduced by the model; \ptopisum is overpredicted at high \pt, 
      whereas \ktopisum is underpredicted in most of the \pt range.
      
\begin{figure}[htb!]
	  \centering
	    \subfloat[\label{fig:dndy_ratio_PPi_0}]{
	    \includegraphics[width=0.35\textwidth]{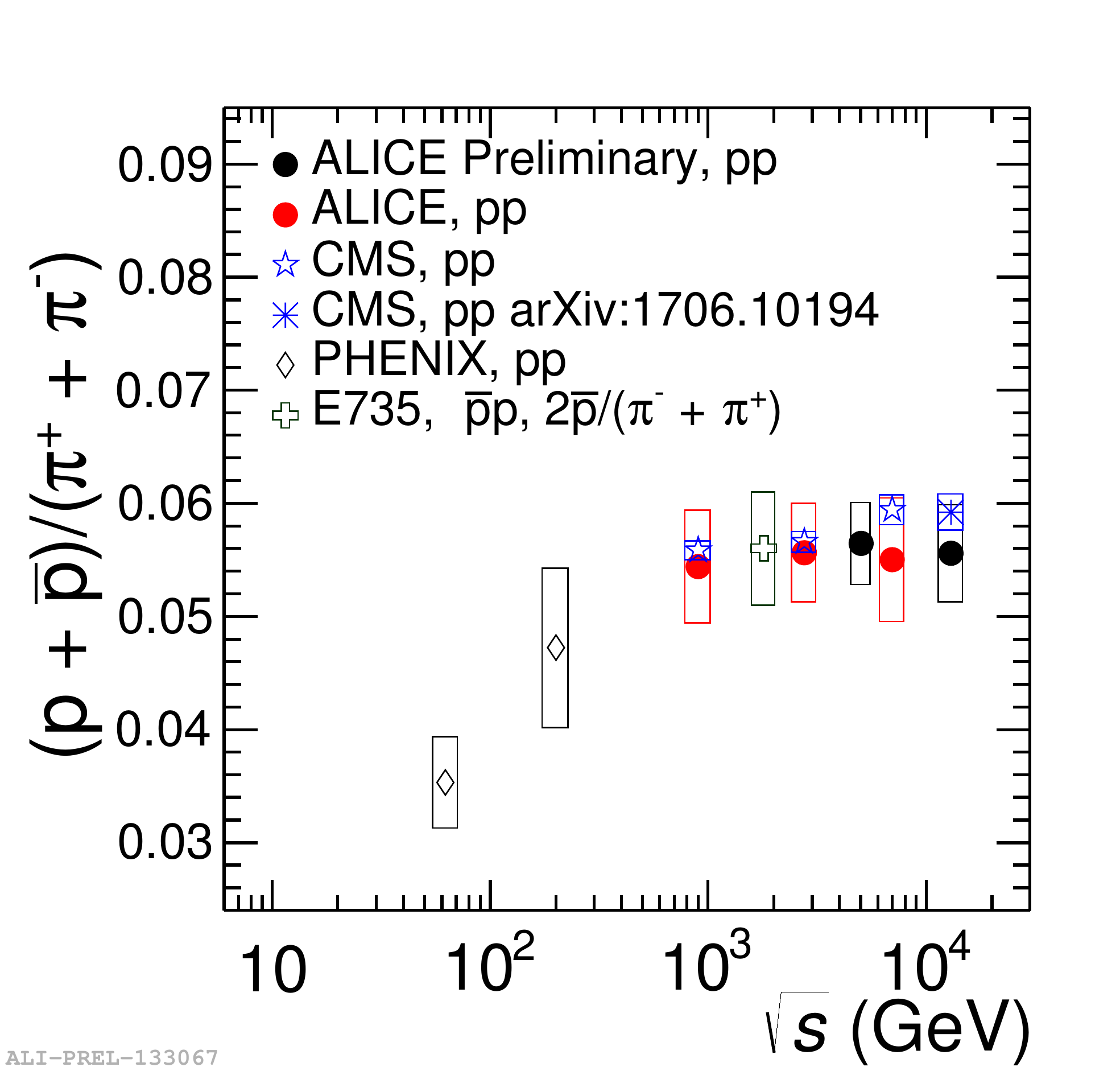}
	    }
	    \subfloat[\label{fig:HyperonToPionRatios}]{
	    \includegraphics[width=0.44\textwidth]{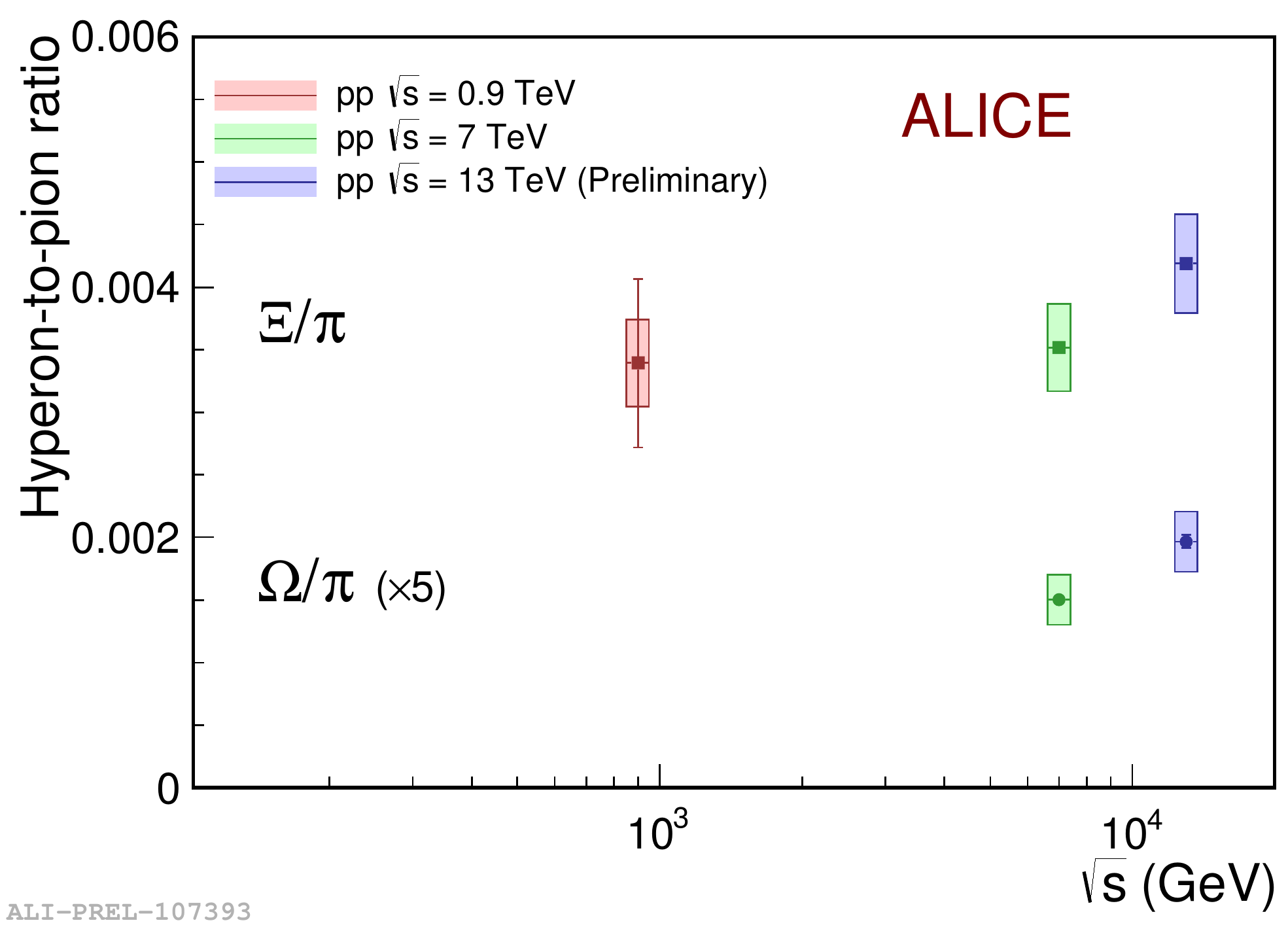}
	    }
	  \caption[]{\dndykrl ratio of \protect\subref{fig:dndy_ratio_PPi_0} \ptopisum and \protect\subref{fig:HyperonToPionRatios} \omegatopisum, and \xitopisum as a function of collision 
	  energy $\sqrt{s}$ (GeV) measured by several experiments in minimum bias \pp (and $\rm{p}\bar{\rm{p}}$) collisions. Open boxes represent the total systematic uncertainties.
	  }
	  \label{fig:dndy_ratio}
	\end{figure}

      Figure~\ref{fig:dndy_ratio} shows the \pt-integrated baryon-to-meson \ptopisum (Fig.~\ref{fig:dndy_ratio_PPi_0}), and hyperon-to-pion \omegatopisum and \xitopisum (Fig.~\ref{fig:HyperonToPionRatios}) ratios as a function of $\sqrt{s}$.
      The \ptopisum ratios with respect to earlier results are complemented with latest measurements for $\sqrt{s}~=~5.02$\,TeV and $\sqrt{s}~=~13$\,TeV \pp collisions, respectively, from ALICE and CMS~\cite{Sirunyan:2017zmn}. While \ptopisum shows saturation in the LHC-energy regime, 
      the hyperon-to-pion ratios exhibit a hint of a modest increase from $\sqrt{s}~=~7$\,TeV to $\sqrt{s}~=~13$\,TeV.
      To further investigate such an effect, one can attempt to test its dependence on multiplicity regardless of collision energy.
      In order to do so, the \pt-integrated yields of \kzs, $\lap+\lam$, $\xim+\xip$, and $\omm+\omp$ have been measured in \pp collisions at $\sqrt{s}=7$ and $\sqrt{s}~=~13$\,TeV as a function of multiplicity~\cite{Vislavicius:2017lfr}.
	Similar particle abundances are observed at similar multiplicities for the two center-of-mass energies. This suggests that particle production and hadrochemistry are dominantly driven by the event 
	activity, represented by the charged-particle multiplicity, rather than by collision energy. \\ 
	Model predictions for \kzs based on the \py~6~\cite{Sjostrand:2006za} and \py~8~\cite{Sjostrand:2007gs} event generators provide a fairly good description, while \eplhc~\cite{Pierog:2013ria} describes the evolution right~\cite{Vislavicius:2017lfr}. In contrast, none of the models capture the evolution 
	of multi-strange hadron yields as a function of $\left <\rm{d}N_{\rm{ch}}/\rm{d}\eta\right >_{|\eta|<0.5}$. Moreover, the discrepancies between data and model predictions become larger for baryons with increasing strangeness content.
        
      Further investigations on particle production as a function of multiplicity showed that the relative yield for particles having non-zero net strangeness content to that of pions (\chpi) increases~\cite{ALICE:2017jyt,Jacazio:2017}. The increase is more 
      pronounced for multi-strange baryons, driven by the strange quark content of the particles alone, regardless of differences in the hadron masses or due to baryon/meson nature of the particle.\\
      The observed increase shows a smooth evolution across different colliding systems, resulting in similar values in high-multiplicity \pp, \ppb, and peripheral \pb collisions at similar multiplicities.
      Among general-purpose MC generators, the \dipsy~\cite{Flensburg:2011kk,Bierlich:2014xba} model describes best the relative increase quantitatively, even though it starts to deviate for triple strange $\Omega$ baryon and at the same time fails to predict 
      the \ptopisum ratio.\\
      None of the Monte Carlo models are able to reproduce all the observations simultaneously.
      Studies of particle production as a function of multiplicity and event shapes have been suggested as a promising tool to attempt to increase the discrimination power for models~\cite{Ortiz:2017jho}.

 %
    
  \section{Conclusions}

  A progressive hardening of the \pt spectral shapes is observed in \pp collisions from $\sqrt{s}=2.76$\,TeV to $\sqrt{s}~=~13$\,TeV. It is shown that the peak position of the \ptopisum ratio experiences a modest 
  shift towards higher \pt with increasing $\sqrt{s}$ which is not reproduced by \py~8.

  The evolution of \pt-integrated yield ratios with $\sqrt{s}$ shows saturation at LHC energies, however there is a hint of a slight increase in the hyperon-to-pion ratios going from $\sqrt{s}=7$ to $\sqrt{s}~=~13$\,TeV.\\
  To factorize the impact of $\sqrt{s}$ , the integrated hadron yields $\mathrm{d}N/\mathrm{d}y$ are measured for single-strange and multi-strange particles at these two distinct center-of-mass energies, as a function of event activity.
  Particle yields show scaling with event activity. Yields are rather similar between different collision energies at comparable multiplicities. This indicates that hadrochemistry is dominantly driven by 
  multiplicity rather than collision energy. The observed scaling behavior is not reproduced by any of the general-purpose MC models.

  One of the most striking observations at the LHC, that is the strangeness enhancement in high-multiplicity \pp and \ppb collisions, is not satisfactorily described by the presently available tunes of the most common MC generators.
  
\section*{Acknowledgments}
\noindent This work was supported by the Hungarian Research Fund (OTKA) under contract No. K120660.

%


\begin{thebibliography}{99}

\bibitem{Voloshin:2008dg} 
  S.~A.~Voloshin, A.~M.~Poskanzer and R.~Snellings,
  arXiv:0809.2949 [nucl-ex].

\bibitem{Bala:2016hlf} 
  R.~Bala, I.~Bautista, J.~Bielcikova and A.~Ortiz,
  Int.\ J.\ Mod.\ Phys.\ E {\bf 25}, no. 07, 1642006 (2016),~doi:10.1142/S0218301316420064
  
\bibitem{ABELEV:2013wsa} 
  B.~B.~Abelev {\it et al.} [ALICE Collaboration],
  Phys.\ Lett.\ B {\bf 726}, 164 (2013),~doi:10.1016/j.physletb.2013.08.024

\bibitem{Khachatryan:2016txc} 
  V.~Khachatryan {\it et al.} [CMS Collaboration],
  Phys.\ Lett.\ B {\bf 765}, 193 (2017),~doi:10.1016/j.physletb.2016.12.009
  
\bibitem{Abelev:2013haa} 
  B.~B.~Abelev {\it et al.} [ALICE Collaboration],
  Phys.\ Lett.\ B {\bf 728}, 25 (2014),~doi:10.1016/j.physletb.2013.11.020
  
\bibitem{Adam:2016dau} 
  J.~Adam {\it et al.} [ALICE Collaboration],
  Phys.\ Lett.\ B {\bf 760}, 720 (2016),~doi:10.1016/j.physletb.2016.07.050
  
\bibitem{ALICE:2017jyt} 
  J.~Adam {\it et al.} [ALICE Collaboration],
  Nature Phys.\  {\bf 13}, 535 (2017),~doi:10.1038/nphys4111
  
\bibitem{Khachatryan:2016yru} 
  V.~Khachatryan {\it et al.} [CMS Collaboration],
  Phys.\ Lett.\ B {\bf 768}, 103 (2017),~doi:10.1016/j.physletb.2017.01.075
  
\bibitem{Rafelski:1982pu} 
  J.~Rafelski and B.~Muller,
  Phys.\ Rev.\ Lett.\  {\bf 48}, 1066 (1982)
  Erratum: [Phys.\ Rev.\ Lett.\  {\bf 56}, 2334 (1986)],~doi:10.1103/PhysRevLett.48.1066
  
\bibitem{Aamodt:2008zz} 
  K.~Aamodt {\it et al.} [ALICE Collaboration],
  JINST {\bf 3}, S08002 (2008),~doi:10.1088/1748-0221/3/08/S08002

\bibitem{Bass:1998vz} 
  S.~A.~Bass, M.~Gyulassy, H.~Stoecker and W.~Greiner,
  J.\ Phys.\ G {\bf 25}, R1 (1999),~doi:10.1088/0954-3899/25/3/013

\bibitem{Abelev:2014ffa} 
  B.~B.~Abelev {\it et al.} [ALICE Collaboration],
  Int.\ J.\ Mod.\ Phys.\ A {\bf 29}, 1430044 (2014),~doi:10.1142/S0217751X14300440
  
\bibitem{Sjostrand:2007gs} 
  T.~Sjostrand, S.~Mrenna and P.~Z.~Skands,
  Comput.\ Phys.\ Commun.\  {\bf 178}, 852 (2008),~doi:10.1016/j.cpc.2008.01.036
  
\bibitem{Brun:1987ma} 
  R.~Brun, F.~Bruyant, M.~Maire, A.~C.~McPherson and P.~Zanarini,
  CERN-DD-EE-84-1.

\bibitem{Adam:2015qaa} 
  J.~Adam {\it et al.} [ALICE Collaboration],
  Eur.\ Phys.\ J.\ C {\bf 75}, no. 5, 226 (2015),~doi:10.1140/epjc/s10052-015-3422-9
  
\bibitem{Adam:2017zbf} 
  J.~Adam {\it et al.} [ALICE Collaboration],
  Phys.\ Rev.\ C {\bf 95}, no. 6, 064606 (2017),~doi:10.1103/PhysRevC.95.064606

\bibitem{Vislavicius:2017lfr} 
  V.~Vislavicius [ALICE Collaboration],
  arXiv:1704.04737 [hep-ex].

\bibitem{Adam:2015pza} 
  J.~Adam {\it et al.} [ALICE Collaboration],
  Phys.\ Lett.\ B {\bf 753}, 319 (2016),~doi:10.1016/j.physletb.2015.12.030
  
\bibitem{Sirunyan:2017zmn} 
  A.~M.~Sirunyan {\it et al.} [CMS Collaboration],
  arXiv:1706.10194 [hep-ex].

\bibitem{Sjostrand:2006za} 
  T.~Sjostrand, S.~Mrenna and P.~Z.~Skands,
  JHEP {\bf 0605}, 026 (2006),~doi:10.1088/1126-6708/2006/05/026
  
\bibitem{Pierog:2013ria} 
  T.~Pierog, I.~Karpenko, J.~M.~Katzy, E.~Yatsenko and K.~Werner,
  Phys.\ Rev.\ C {\bf 92}, no. 3, 034906 (2015),~doi:10.1103/PhysRevC.92.034906
  
\bibitem{Jacazio:2017} 
  N.~Jacazio [ALICE Collaboration],
  J.\ Phys.\ Conf.\ Ser.\  {\bf 832}, no. 1, 012019 (2017),~doi:10.1088/1742-6596/832/1/012019
  
\bibitem{Flensburg:2011kk} 
  C.~Flensburg, G.~Gustafson and L.~Lonnblad,
  JHEP {\bf 1108}, 103 (2011),~doi:10.1007/JHEP08(2011)103
  
\bibitem{Bierlich:2014xba} 
  C.~Bierlich, G.~Gustafson, L.~L\"onnblad and A.~Tarasov,
  JHEP {\bf 1503}, 148 (2015),~doi:10.1007/JHEP03(2015)148
  
\bibitem{Ortiz:2017jho} 
  A.~Ortiz,
  arXiv:1705.02056 [hep-ex].
  
\end{thebibliography}
\end{document}